\documentclass[11pt]{article}
\usepackage{amssymb}
\usepackage[dvips]{graphicx}

\def\a{\alpha}

\def\d{\delta}
\def\D{\Delta}

\def\e{\epsilon}

\def\l{\lambda}

\def\m{\mu}
\def\n{\nu}
\def\s{\sigma}
\def\S{\Sigma}

\def\r{\rho}

\newcommand{\sm}[1]{\mbox{\scriptsize #1}}

\renewcommand{\@}[1]{\sqrt{#1}}

\renewcommand{\le}[1]{\label{#1}\end{eqnarray}}
\newcommand{\be}{\begin{equation}}
\newcommand{\ee}{\end{equation}}
\newcommand{\bea}{\begin{eqnarray}}
\newcommand{\eea}{\end{eqnarray}}
\newcommand{\nn}{\nonumber}

\newcommand{\eq}[1]{(\ref{#1})}
\def\nn{\nonumber\\}

\def\ffract#1#2{\raise .35 em\hbox{$\scriptstyle#1$}\kern-.25em/
\kern-.2em\lower .22 em \hbox{$\scriptstyle#2$}}

\def\half{{1\over2}\,}
\begin{document}

\pagestyle{empty}

\begin{flushright}
AEI-2005-146\\
{\tt hep-th/0509167}\\
\end{flushright}
\vskip0.1truecm

\begin{center}
\vskip 3truecm {\Large \textbf{A Note on Knot Invariants and $q$-Deformed 2d Yang-Mills}}

\vskip 1.5truecm

{\large \textbf{Sebastian de Haro}}\\
\vskip .9truecm {\it Max-Planck-Institut f\"{u}r
Gravitationsphysik\\
Albert-Einstein-Institut\\
14476 Golm, Germany}\\
\tt{sdh@aei.mpg.de} 

\end{center}

\vskip 4truecm

\begin{center}
\textbf{\large Abstract}
\end{center}

We compute expectation values of Wilson loops in $q$-deformed 2d Yang-Mills on a Riemann surface and show that 
they give invariants of knots in 3-manifolds which are circle bundles over the Riemann surface. The
areas of the loops play an essential role in encoding topological information about the extra dimension, 
and they are quantized to integer or half integer values.



\newpage

\pagestyle{plain}

\section{Introduction}

Recently, there has been a lot of interest in $q$-deformed 2d Yang-Mills theory \cite{aosv,sdh2,sdhmt2,nv,abms,
jm,ccgpss}. This theory realizes the idea of Gross and Taylor \cite{gross,grosstaylor,vafa} in a very concrete 
way. It describes closed topological A-model strings with bound states of D4, D2, and D0-branes in Calabi-Yau spaces which 
are two complex line bundles over a Riemann surface. This Riemann surface is the Riemann surface where $q$-2d Yang-Mills
theory lives, and the $q$-2d Yang-Mills action is the dimensionally reduced action on the D4-brane worldvolume.
It also has an interesting 
Douglas-Kazakov type phase transition \cite{abms,jm,ccgpss}. On the open string side, we have Chern-Simons 
theory as the effective target space field theory description. 

So it is not unreasonable to expect that, as a result of geometric
transition, $q$-2dYM observables somehow 
produce topological invariants of 3-manifolds and invariants of knots embedded in these three-manifolds. This
idea goes back to \cite{witten2d,polyakreshetikhin}. It is known \cite{sdh1,aosv,sdh2} that the partition function of $q$-2dYM on
a Riemann surface $\Sigma$ reproduces the partition function of Chern-Simons on a Seifert space, that is a 
3-manifold $M$ that is the total space of a circle bundle over this Riemann surface. 
The partition function of Chern-Simons in Seifert spaces was recently 
analyzed in \cite{beasleywitten} using non-abelian localization. 

The question we will answer in this letter is whether this is still true for Wilson loops in $q$-2dYM, that is,
whether Wilson loops still produce topological invariants of Seifert spaces, and whether the invariant
is the Reshetikhin-Turaev invariant of knots in 3-space. At a first glance, one would think that this is impossible. 2d 
Yang-Mills theories are invariant under area preserving diffeomorphisms, and in particular Wilson loops have 
an exponential decay as a function of the area of the loop. So naively one would think that such information 
is not present in Chern-Simons theory at all. The resolution is that the areas are quantized and actually
encode topological information concerning upper/under passes of the knot. We will see that Wilson loops 
indeed give knot invariants, {\it provided} the areas are quantized to half-integer values. There is an 
$H_1(\S)$ ambiguity in reconstructing invariants of $M$ from the Wilson loop in $q$-2dYM. Full details 
will appear in a separate paper \cite{sdh}.

Let us emphasize that all our results are at finite $k$ and $N$. We list here some of the questions which 
we will {\it not} address in this paper. First of all, we will limit 
ourselves to cases where the Wilson loops do not have crossings. Nevertheless, they give invariants of knots 
with crossings. This is possible because of a two-dimensional version of the Reidemeister
moves which allows us to remove certain crossings on the surface. Cases where these crossings cannot be 
removed will be considered in \cite{sdh}. Also, let us notice that $q$-deformed 2dYM was originally defined in 
\cite{buffenoirroche} using quantum groups. One actually expects the quantum characters to play a role in the
prescription of \cite{aosv}. Indeed, this is needed in order for the plaquette gluing properties to be consistent
with quantum (rather than classical) dimensions\footnote{This remark is further developed in \cite{sdhsr}.}.
Finally, we only consider the case of 
$q$ being a root of unity here, but it should be possible to generalize our results to other values of $q$.

As part of our motivation, we should also mention the fact that 2d Yang-Mills theories have interesting
phenomenological applications \cite{blnt,ambjorn}. For small Wilson loops they describe the confining phase of 
QCD to a good degree of accuracy.

\section{$q$-deformed 2d Yang Mills without Wilson lines}

The partition function of $q$-deformed 2dYM on a closed Riemann surface of genus $g$ is given by \cite{aosv}:
\be\label{partfunction}
Z_{\sm{q2dYM}}(\S_g)=\sum_\l(\mbox{dim}_q(\l))^{2-2g}\,q^{{p\over2}\,C_2(\l)}~.
\ee
Since we are working at $q$ root of unity, $q=e^{2\pi i\over k+N}$ where $k$ is the Chern-Simons coupling,
the sum runs over the integrable representations of the gauge group $G$ only, $P_+$. For $G=SU(2)$, this 
means that the spin is bounded by $j\leq k$. If $q$ is not a root of unity, the sum is over all 
irreducible representations. The quantum dimensions are given by
\be
\mbox{dim}_q(\l)=\prod_{\a>0}{([\l+\r,\a)]\over[(\r,\a)]}=\prod_{\a>0}{\sin{\pi(\l+\r,\a)\over k+g}
\over\sin{\pi(\r,\a)\over k+g}}
\ee
where $\r$ is the Weyl vector which labels the trivial representation, $g$ is the dual coxeter number, and 
the inner product is taken with respect
to the Cartan metric (for our group theory conventions, see appendix \ref{liea}). For $G=SU(N)$, this is 
simply
\be
\mbox{dim}_q(\l)=\prod_{1\leq j<i\leq N}{\sin{\pi(\ell_i-\ell_j+j-i)\over k+N}\over\sin{\pi(j-i)\over k+N}}
\ee
where $\ell_i$ is the number of boxes in the $i$th row of the Young tableau. The $q$-numbers are defined as usual,
\be\label{qdim}
[x]={q^{x/2}-q^{-x/2}\over q^{1/2}-q^{-1/2}}~.
\ee

It is easy to see that \eq{partfunction} is the partition function of Chern-Simons in a Seifert space which
is a circle fibration over $\S_g$ with Chern class $p$ \cite{sdh1,sdhmt1}. In the rest of the paper we will
denote this space $M$. It can be obtained by
surgery on $S^1\times\Sigma_g$ with a gluing operation $U=ST^pS$. In the case $\S_g=S^2$, this just gives the 
lens space $S^3/\mathbb{Z}_p$. Remember that $S$ and $T$ are the $PSL(2,\mathbb{Z})$ generators,
\bea\label{sl2Z}
S^2&=&C\nn
(ST)^3&=&C
\eea
where $C$ is the charge conjugation matrix, and they are represented on affine characters \cite{jones}, so
matrix indices label integrable representations of $G$, $S_{\l\m}$ and $T_{\l\m}$. $T$ is actually
diagonal (see \eq{modularmatrices} in Appendix \ref{liea}) so we will denote it by $T_\l$. It is easy to check
that
\be
{S_{\l\r}\over S_{\r\r}}=\mbox{dim}_q\l~.
\ee

Now rewriting \eq{partfunction} in terms of $S$ and $T$ using the above and the definition of $T$ in Appendix \ref{liea},
it is clear that \eq{partfunction} is the operator $ST^pS$ evaluated on the trivial representation,
\be
Z_{\sm{q2dYM}}[\S_g;p]={1\over S_{\r\r}^2}\,(ST^pS)_{\r\r}~,
\ee
hence it equals the partition function of Chern-Simons theory in $M$. The normalization is conventional and due to
our choice \eq{qdim}. The case $p=1$ is of course just the 
three-sphere; applying the defining relation \eq{sl2Z} $STS=CT^{-1}S^{-1}T^{-1}$, we just get
\be
Z_{\sm{q2dYM}}[S^2,p=1]=T_{\r\r}^{-2}S_{\r\r}
\ee
which is indeed the partition function on $S^3$ \cite{jones} (in non-canonical framing).

\section{Wilson loops in $q$-2d Yang Mills}

$q$-deformed 2d Yang-Mills is most naturally defined in terms of quantum groups \cite{buffenoirroche}\footnote{See also \cite{citrad}.}.
In this
paper, we will work at the level of the $q$-2d Yang Mills amplitudes, leaving off-shell questions for the 
future. Let us just mention that the procedure to compute Wilson loops in $q$-2dYM is 
similar to the underformed case \cite{cmr}. We obtain it by Migdal's cut-and-paste procedure, but now this is
much more subtle since the variables are quantum group elements, more precisely they live in the quantum enveloping
algebra of the gauge group. Nevertheless character integration formulas are known for the quantum case, and they 
are expressed in terms of quantum 3j- and 6j-symbols \cite{kr} -- the $q$-deformed analogs of the quantities familiar
from quantum mechanics.

\begin{figure}
\begin{center}
\includegraphics
{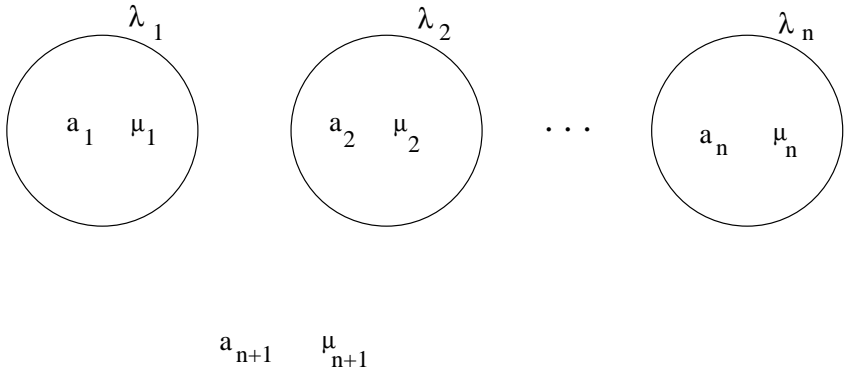}
\caption{\small $n$ Wilson loops in representations $\l_1,\ldots,\l_n$ with areas $a_1,\ldots,a_n$. The outer face
has area $a_{n+1}$ and Euler number $2-2g-n$. $\m_1,\ldots\m_{n+1}$ are the representations of the faces.}
\label{circles}
\end{center}
\end{figure}

In this section we mostly consider the case of no intersections between the loops on the Riemann surface. This means 
that we will only need 3j-symbols. These are precisely the fusion coefficients in conformal field theory. So the
expectation value of $n$ Wilson loops in representations $\l_1\ldots\l_n$ with areas $a_1,\ldots,a_n$ (see in 
Figure \ref{circles}) is given by
\be\label{wilsonloops}
Z_{\sm{q2dYM}}(\Sigma_g,a_i,\l_i)=\sum_{\m_1\ldots\m_{n+1}}\prod_{i=1}^{n+1}(\mbox{dim}_q(\m_i))^{\chi(\m_i)}q^{-{a_i\over2}\,C_2(\m_i)}
\prod_{j=1}^nN^{(k)\l_j}_{\m_j\m_{n+1}}
\ee
where $N^{(k)}$ are the fusion coefficients in the WZW model at level $k$. The sum is over the integrable representations
of $G$. From now on we will always assume that we work at level $k$ and drop the superscript $(k)$.

In the following we will consider the case $\chi(\m_i)=1$ for $i=1,\ldots,n$, $\chi(\m_{n+1})=2-2g-n$.
A further (minor) restriction in writing \eq{wilsonloops} is that we specialized to non-concentric Wilson loops, but the 
formula in this case is almost identical to \eq{wilsonloops}.

The case of intersecting Wilson loops will be discussed in \cite{sdh} in full generality.

In what follows we work out the formula \eq{wilsonloops} and show how it reproduces various interesting knot invariants
in the Seifert fibered space. The invariants crucially depend on the values of the areas, $a_1,\ldots,a_n$, which indeed
contain most of the topological information about the three-dimensional Chern-Simons theory.

In order to do this, it will be useful to use the Verlinde formula for the fusion coefficients:
\be
N_{\l\m}^\n=\sum_\s{S_{\l\s}S_{\m\s}S^*_{\n\s}\over S_{\s\r}}
\ee
where as usual $\r$ denotes the trivial representation. This formula allows us to perform most of the sums
in the partition function \eq{wilsonloops}. From it we can prove the following formulas (see appendix 
\ref{liea}):
\bea\label{fusioneqs}
\sum_\l\mbox{dim}_q(\l)\,N_{\m\n}^\l&=&\mbox{dim}_q(\m)\,\mbox{dim}_q(\n)\nn
\sum_\l\mbox{dim}_q(\l)\,N_{\m\l}^\n\, T_\l &=&{T_{\r\r}^{-1}\over S_{\r\r}}\,(TST)_{\m\n}
\eea

We now work out some examples of \eq{wilsonloops} and show how different choices of the areas $a_1,\ldots,a_{n+1}$ give different knot 
invariants in $M$. Our choice of areas is motivated by Turaev's shadow invariant \cite{turaev} and will be explained in \cite{sdh}.
The total area of the surface equals the Chern class of the fibration,
\be
\sum_{i=1}^{n+1}a_i=p~,
\ee
in agreement with \cite{aosv}.

\subsection{Chern-Simons on $S^1\times\Sigma$}

In this subsection we consider Chern-Simons on the trivial fibration $S^1\times\S$, that is we take $p=0$.

\subsubsection{Zero winding}

The case $p=0$ and $n$ unknots lying on $S^2$ with zero winding around the $S^1$ corresponds to $n$ disjoint Wilson loops on the
$q$-2dYM side, in the limit of vanishing areas (see Figure \ref{circles}), so $a_i=0$ for $i=1,\ldots,n+1$. We get:
\be
W^{(0,0,n)}_{\l_1\ldots \l_n}=\sum_{\mu_1\ldots\mu_{n+1}}\mbox{dim}_q(\m_1)\ldots\mbox{dim}_q(\m_n)(\mbox{dim}_q(\m_{n+1}))^{2-n}
N_{\m_1\m_{n+1}}^{\l_1}N_{\m_2\m_{n+1}}^{\l_2}\ldots N_{\m_n\m_{n+1}}^{\l_n}~.
\ee
where the triple $(p,g,n)$ labels the Chern class of the bundle, the genus of $\Sigma$ and the number of loops, respectively.
Clearly, the summations can be performed using the first of \eq{fusioneqs}. We get:
\be
W_{\l_1\ldots\l_n}^{(0,0,n)}={S_{\l_1\r}\ldots S_{\l_n\r}\over S_{\r\r}^n}~,
\ee
This is precisely the Chern-Simons result, where the product of the $n$ unlinked loops factorizes in this case.

This is now easy to generalize to a surface of genus $g$. The only change is that the Euler number of the last face is now $2-2g-n$, therefore
we are left with:
\be
W_{\l_1\ldots\l_n}^{(0,g,n)}={S_{\l_1\r}\ldots S_{\l_n\r}\over S_{\r\r}^n}\,\sum_\l(\mbox{dim}_q(\l))^{2-2g}~.
\ee
Obviously, setting $n=0$ we just get the partition function of Chern-Simons on $S^1\times\Sigma$:
\be
Z_{\sm{CS}}(S^1\times\Sigma)=\sum_\l(\mbox{dim}_q(\l))^{2-2g}~.
\ee
This is the well-known Verlinde formula for the dimension of the space of conformal blocks on a surface of genus $g$. It is obtained from the
partition function of the $q$-deformed 2dYM on $\Sigma$.

\subsubsection{Non-zero winding}

Let us now consider some less trivial examples involving winding around the $S^1$. We want to work out the expectation values of $n$ unknotted,
unlinked circles winding the $S^1$ with winding number 1. 

Let us work out a single loop first, that is $n=1$ and take $\Sigma=S^2$. This now corresponds to the same Figure \ref{circles}
with $n=1$, $a_1=1$ and $a_2=-1$. Thus, we get
\be
W_\l=\sum_{\m\n}\mbox{dim}_q(\m)\,\mbox{dim}_q(\n)\,N_{\m\n}^\l\,T_\m T_\n^{-1}~.
\ee
Using the second of \eq{fusioneqs}, we get precisely
\be
W_\l^{(0,0,1;1)}={1\over S_{\r\r}^2}\,\d_{\l\r}
\ee
where we included an additional label $(p,g,n;w)$ for the winding. This agrees with the fact that the Hilbert space
on $S^1\times S^2$ with one marked point on the $S^2$ has dimension 1 if $\l$ is trivial, and zero otherwise \cite{jones}.

We can now do the case of $n$ loops with winding number $1$ around the $S^1$. In this case we have Figure \ref{circles} with 
$a_i=1$ for $i=1,\ldots,n$, and $a_{n+1}=-n$ for the outer face. We get:
\be
W_{\l_1\ldots\l_n}^{(0,0,n;1)}={T_{\l_1}\ldots T_{\l_n}\over T_{\r\r}^n S_{\r\r}^2}\,\sum_\m S_{\m\r}^{2-n}S_{\l_1\m}\ldots S_{\l_n\m}
\ee
In particular, in the cases $n=2$ and $n=3$ we get
\bea
W_{\l\m}^{(0,0,2;1)}&=&{T_\l T_\m\over T_{\r\r}^2S_{\r\r}^2}\,\d_{\l\m^*}\nn
W_{\l\m\n}^{(0,0,3;1)}&=&{T_\l T_\m T_\n\over T_{\r\r}^3S_{\r\r}^2}\,N_{\l\m\n}~,
\eea
in agreement with \cite{jones} for the dimensions of the Hilbert space of conformal blocks on the sphere with two and three marked
points, respectively. Notice however that on the $q$-2d Yang-Mills side the loops are completely regular rather than being marked points.
This is due to the way the projection is done. The framing comes out differently 
above due to the winding around the $S^1$. To reach canonical framing we need to untwist every loop by a factor of $T_\l^{-1} T_{\r\r}$.
We will ignore framing ambiguities in the rest of the paper. The additional factors of $S_{\r\r}$ are due to our choice of normalization 
of the quantum dimensions \eq{qdim}. 

We can easily generalize this to the case of a Riemann surface $\S$. The areas are the same as before, the only difference being the
Euler number:
\be
W_{\l_1\ldots\l_n}^{(0,g,n;1)}={T_{\l_1}\ldots T_{\l_n}\over T_{\r\r}^n }\sum_\l(\mbox{dim}_q(\l))^{2-2g-n}\,S_{\l_1\r}\ldots S_{\l_n\r}~.
\ee
This is the Verlinde formula for the dimension of the space of conformal blocks on a Riemann surface of genus $g$ with $n$ punctures, labeled 
by representations $\l_1\ldots\l_n$ \cite{blauthompson}. Indeed, in canonical quantization of Chern-Simons the above corresponds to $n$ Wilson 
lines that pierce the Riemann surface \cite{jones}. In $q$-2dYM we see that they are computed by $n$ regular Wilson loops
lying on the Riemann surface as in Figure \ref{circles}, but now with non-zero area.

As a last example on the trivial bundle, we compute the invariant of a single loop with winding $w$ around the $S^1$. The areas are 
$a_1=-a_2=w$. We get:
\be
W_\l^{(0,0,1;w)}=\sum_{\m\n}\mbox{dim}_q(\m)\,\mbox{dim}_q(\n)\,N_{\m\n}{}^\l\,T_\m^w T_\n^{-w}~.
\ee
Filling in the fusion coefficients, we get
\be
W_\l^{(0,0,1;w)}={1\over S_{\r\r}^2}\sum_\m U^{(-1,w)}_{\r\m}U^{(-1,-w)}_{\m\r}\,S_{\l\m}\,{1\over S_{\m\r}}
\ee
where $U^{(-1,w)}=ST^wS$ (in the notation of \cite{sdh1}) is the $SL(2,\mathbb{Z})$ operator that glues two solid tori into a lens space, 
as discussed in section 2.

\subsection{Chern-Simons in Seifert fibered spaces}

Having done $S^1\times \Sigma$, non-trivial bundles can be easily described in $q$-2dYM: we simply increase the area of the outer
face by an amount $p$ equal to the Chern class of the bundle!

Before doing this, let us recall how to get the Chern-Simons expressions \cite{jones,jeffrey,rozansky,marcos}. Consider performing surgery
on $S^1\times \Sigma$ with an operator $U\in SL(2,\mathbb{Z})$ to obtain a Seifert fibered manifold $M$. We get a link $L(C_1,\ldots,C_n)$
in $M$ if we start with a link in $S^1\times\S$. For example, take $n$ unknots around the $S^1$ and apply the modular
transformation $U$:
\be
Z_{\sm{CS}}[M,L(C_1,C_2,\ldots,C_n;\l_1,\ldots,\l_n)]=\sum_\l U_{\l_n}^\l\,Z_{\sm{CS}}[S^1\times\S;C_1,\ldots,C_n;\l_1,\ldots,\l_{n-1},\l]
\ee
where $\l_1,\ldots,\l_n$ are the representations of the components of the link $L(C_1,\ldots,C_n)$ (we are raising and lowering indices with
the charge conjugation matrix). Thus, from the formulas in the previous section, we have
\be
Z_{\sm{CS}}[M,L(C_1,C_2,\ldots,C_n;\l_1,\ldots,\l_n)]=\sum_{\l\m} U_{\l_n}^\l S_{\m\r}^{2-2g-n}S_{\l_1\m}\ldots S_{\l_{n-1}\m}S_{\l\m}~.
\ee
As a simple example, take $U=S$ so that $M=S^3$. Using the defining relations \eq{sl2Z}, we trivially get 
\bea\label{S3}
Z_{\sm{CS}}[S^3;L(\l,\m)]&=&S_{\l\m}\nn
Z_{\sm{CS}}[S^3;L(\l,\m,\n)]&=&\sum_\s S_{\n\s} N_{\l\m\s}~.
\eea
The first equation is the result for the Hopf link; the second one corresponds to two paralel, unlinked circles with a third circle 
linked with them with link number one for each unknot. In \cite{jones}, this gave a new proof of the Verlinde formula.

For $M$ a circle fibered Seifert manifold over $\S$, $U=(ST^pS)$ and we get for instance
\bea\label{seifert}
Z_{\sm{CS}}[S^3/\mathbb{Z}_p]&=&(ST^pS)_{\r\r}\nn
Z_{\sm{CS}}[S^3/\mathbb{Z}_p;L(\l,\m)]&=&(ST^pS)_{\l\m}\nn
Z_{\sm{CS}}[M;L(\l_1,\ldots,\l_n)]&=&\sum_{\l\m}(ST^pS)_{\l_n}^\m S^{2-2g-n}_{\l\r}S_{\l_1\l}\ldots S_{\l_{n-1}\l}S_{\m\l}\nn
&=&\sum_\l T^p_\l S^{2-2g-n}_{\l\r}S_{\l_1\l}\ldots S_{\l_n\l}~.
\eea

As an example, we will now reproduce these formulas from $q$-2d Yang Mills.
We first work out the case of zero winding. We now take $a_i=0$, $i=1,\ldots,n$ and $a_n=p$ in Figure \ref{circles}. We get:
\be
W_{\l_1\ldots\l_n}^{(p,g,n;0)}=S_{\l_1\r}\ldots S_{\l_n\r}\sum_\l (\mbox{dim}_q(\l))^{2-2g}T^p_\l~.
\ee
In the case $p=n=1$, $g=0$, after using the defining properties of $S$ and $T$ we get:
\be\label{1010}
W_{\l}^{(1,0,1;0)}={T_{\r\r}^{-1}\over S_{\r\r}^3}\,S_{\l\r}~.
\ee
reproducing the unknot in $S^3$ \eq{S3} -- the vertical and the horizontal loop give homological circles in $S^3$. 
For several unlinked unknots, we get
\be
W_{\l_1\ldots\l_n}^{(1,0,n;0)}={T_{\r\r}^{-1}\over S_{\r\r}^3}\,S_{\l_1\r}\ldots S_{\l_n\r}~,
\ee
indeed the disjoint product of $n$ unknots.

In the case of $n$ unknots with winding $w=1$ around the $S^1$, we have $a_i=1$ for $i=1,\ldots,n$ and $a_{n+1}=p-n$. We get
\be
W_{\l_1\ldots\l_n}^{(p,g,n;1)}={T_{\l_1}\ldots T_{\l_n}\over T_{\r\r}^n}\sum_\l(\mbox{dim}_q(\l))^{2-2g-n}\,
T_\l^p S_{\l_1\l}\ldots S_{\l_n\l}~.
\ee
This is the general formula for the link $L(\l_1,\ldots,\l_n)$ in the Seifert manifold $M$ considered in \eq{seifert}, and the 
most general example we consider in this paper.

Of course, we get the partition function of Chern-Simons in the $S^1$ bundle over $\Sigma$ if we take $n=0$:
\be
W^{(p,g,0)}=\sum_\l(\mbox{dim}(\l))^{2-2g}T_\l^p=S_{\r\r}^{2g-2}\,Z_{\sm{CS}}(M)~.
\ee

\subsection{A comment on crossings points}

We already mentioned that in order to properly deal with crossings in $q$-2d Yang Mills, we need to include quantum 6j-symbols.
Nevertheless, most of the examples above did contain crossings, so what happened? 

It turns out that there are Reidemeister moves in two
dimensions that allow us to remove some of these crossings. These moves depend on the area's. Presumably the examples above are all cases 
where such moves are applicable. 
We will come back to this issue in the future.

\newpage

\section*{Conclusions and future directions}

We have shown that $q$-deformed 2d Yang-Mills theory on a Riemann surface $\S$ gives topological invariants in one dimension higher,
namely in a Seifert fibered manifold. More precisely, expectation values of Wilson loops on the Riemann surface give knot invariants
of the Seifert manifold. The
areas of the Wilson loops of the surface play an essential role in encoding topological information about the extra dimension; they are
quantized in integer or half-integer values (in this paper we considered integer values only). In fact, $q$-2d Yang Mills gives a very
effective way to compute knot invariants. In many cases, the invariants can be computed without taking into account 6j-symbols at the
crossings. This is due to an underlying analog of the Redemeister moves in two dimensions. The infinite number of choices of (integer
or half-integer) areas
of Wilson loops generate an infinite amount of knot invariants in three dimensions. Our choice of areas was motivated by Turaev's
{\it shadow invariant} \cite{turaev}. We will develop this more general point of view in a future publication. A path integral derivation
of the shadow invariant for the case $S^1\times \S$ is given in \cite{sdhah}.

In some of our formulas we needed to take the areas to negative values for comparison with relatively simple Chern-Simons observables; it 
would be interesting to see whether from the brane point of view \cite{aosv} these configurations are pathological or not. In any case, 
the case where all areas are positive also provides knot invariants in the Seifert manifold, be it not the simplest ones. Notice that the 
total area always equals $p$.

It would be interesting to study open-closed string duality and geometric transitions with non-trivial brane configurations in the 
topological string from the point of view developed in this paper. In fact, that was our original motivation. This will correspond to 
the A-model amplitudes with branes inserted. In this case one needs to
generalize the formulas in this paper to non-root-of-unity values of $q$. This would then be similar in spirit to \cite{ani,okuda}, where 
certain Chern-Simons invariants are obtained from the crystal prescription. We hope to come back to these issues in the future.

\section*{Acknowledgements}

We thank Atle Hahn, Sanjaye Ramgoolam, and Alessandro Torrielli for colaboration in closely related topics. We also thank 
Nicolai Reshetikhin, Ingo Runkel, Volker Schomerus, Stefan Theisen, Miguel Tierz, and Vladimir Turaev for discussions. We thank the organizers of the 
Amsterdam Summer Workshop and the Simons Workshop in Mathematics and Physics 2005, where part of this work was done.

\appendix

\section{Modular matrices and Lie algebra conventions}\label{liea}

In this appendix we summarize some of the Lie group conventions we used in the
main text, provide the explicit expression for the modular matrices, and prove
formula \eq{fusioneqs}.

We first work out some formulas for $U(N)$. The relation between the number of boxes 
in the Young tablaux and the weight of a representation in the usual fundamental weight
basis is:
\be
\ell_i=\l_i+\l_{i+1}+\ldots\l_N~.
\ee
The Weyl vector labeling the trivial representation is:
\be
\r=\sum_{i=1}^N\left({N+1\over2}-i\right)\e_i
\ee
where $e_i$ is a unit vector in $\mathbb{R}^N$. Its norm is
\be
|\r|^2={1\over12}\,N(N^2-1)~.
\ee
The second Casimir becomes especially simple in following basis:
\be
h_i=\ell_i+\r_i~.
\ee
It is:
\bea
C_2(\l)&=&(\l,\l+2\r)\nn
&=&|\ell+\r|^2-|\r|^2=|h|^2-|\r|^2=\sum_{i=1}^Nh_i^2-{1\over12}\,N(N^2-1)~.
\eea

The $SL(2,\mathbb{Z})$ generators $S$ and $T$ used in the main text are given by:
\bea\label{modularmatrices}
T_{\l\m} &=&\delta _{\l\mu}\,e^{{2\pi iC(\l)\over2(k+g)}-{2\pi ic\over24}}\nn
S_{\l\m} &=&{i^{|\D_+|}\over(k+g)^{r/2}}\,|P/Q^\vee|^{-\half}
\sum_{w\in W}\epsilon (w)e^{-{2\pi i\over k+g}(\l,w\cdot\m)}~,
\eea
where the central charge is $c=k\,{\mbox{dim}}\,g/(k+g)$. For explicit determinantal formulas in the
case of $U(N)$, $SO(N)$ and $Sp(N)$, see \cite{sdh2}.
In the main text we dropped the overall normalization of $T$.

The proof of
\be
\sum_\l\mbox{dim}_q(\l)N^\m_{\l\n}={S_{\m\r}S_{\n\r}\over S_{\r\r}^2}
\ee
is starightforward. We fill in the Verlinde formula for $N^\m_{\l\n}$ and use $S^2=C$. The proof of
\be
\sum_\l\mbox{dim}_q(\l)\,N^\m_{\l\n}\,T_\l={T_{\r\r}^{-1}\over S_{\r\r}}(TST)_{\m\n}
\ee
is similar. We fill in the Verlinde formula for $N^\m_{\l\n}$ and use $(ST)^3=C$ twice. The result follows.

\end{document}